\documentstyle[sprocl]{article}

\input{epsf}

\def\be{\begin{equation}}
\def\ee{\end{equation}}
\def\bea{\begin{eqnarray}}
\def\eea{\end{eqnarray}}

\begin{document}

\title{The non-perturbative constraint on sea quarks\\
---the strange sea quarks in the nucleon
and the soft pion contribution at high energy---}

\author{Susumu Koretune}

\address{Department of Physics, Shimane Medical \vspace{-1mm}University\\
Izumo, Shimane, \vspace{-1mm}Japan\\E-mail: koretune@shimane-med.ac.jp}


\maketitle\abstracts{The mean charge sum rule for the light sea quarks 
in the nucleon which holds under the same theoretical footing 
as the modified Gottfried sum rule shows that
a usual parameterization of the strange sea quark distribution
underestimate its contribution in the small $x$ region.
We give a discussion of the soft pion contribution at high energy 
as a possible explanation of the saturation of the sum rule and show 
that it naturally explains why the strange sea quark is suppressed
in the region above $x \sim 0.01$ while it becomes abundant below it.}
\section{Introduction}
The current anticommutation relation on the null plane
based on the canonical quantization and
the Deser, Gilbert, and Sudarshan (DGS) representation~\cite{DGS} was given
many years ago.~\cite{kore80}
The obtained relation has not been the operator relation but
the connected matrix element between the one particle stable hadron.
This is because the spectral condition plays an essential role to derive
the relations. In spite of this limitation, however, we have obtained many sum rules
due to the chiral $SU(N)\otimes SU(N)$ flavor symmetry on the null-plane.
These sum rules have given us the non-perturbative constraint on the sea quark
distributions in the hadrons.
The modified Gottfried sum rule is one example
and has explained the experimental value of the Gottfried sum 
model independently.~\cite{kore93} In this talk, I give the study of the
strange sea quark distribution based on the mean charge sum rule 
for the light sea quarks in the nucleon,  
and give a discussion of the soft pion contribution at high energy 
as a possible explanation of the saturation of the sum rule. 
\section{Sum rules}
From the current anticommutation relation on the null-plane,
the modified Gottfried sum rule has been obtained as
\begin{eqnarray}
\lefteqn{\int^1_0\frac{dx}{x}\{F_2^{ep}(x,Q^2)-F_2^{en}(x,Q^2)\}}
\nonumber \\
 &=&\frac{1}{3}\left( 1-\frac{4f_{K}^2}{\pi}\int_{m_Km_N}^{\infty}\frac{d\nu}{\nu^2}
\sqrt{\nu^2-(m_Km_N)^2}\{\sigma^{K^+n}(\nu)-\sigma^{K^+p}(\nu)\}\right) .
\end{eqnarray}
Based on this sum rule the Gottfried sum has been predicted as
$0.26\pm 0.03$. A numerical evaluation of this sum rule 
suggests an importance of the contributions 
from high energy regions to explain the departure from $1/3$.~\cite{kore93}
Further, from a theoretical point of view, the high energy and the low energy
is heavily related in the sum rule.~\cite{weinberg}\\
Now, by studying the structure of the sum rule we see that $3/2$ of this quantity 
corresponds to [(the mean $I_3$ of quarks) - (the mean $I_3$ of anti-quarks)].
We can check this facts by multiplying the left-hand side of the sum rule by $3/2$,
and express it by the parton model as
$\int_0^1dx\{\frac{1}{2}u_v - \frac{1}{2}d_v\} +
\int_0^1dx\{\frac{1}{2}\lambda_u - \frac{1}{2}\lambda_d\}
- \int_0^1dx\{- \frac{1}{2}\lambda_{\bar{u}} + 
\frac{1}{2}\lambda_{\bar{d}}\} $
where $\lambda_{i}$ means the sea quark of the $i$ type.
Thus we see that the sum rules obtained by the current anticommutation 
relation give us information of the mean quantum number of the sea quarks in the hadron.
Here we must be careful about the following facts when we consider the phenomenological
analysis. The sea quark distribution functions in our analysis are 
defined only as the integrated quantity through the moment at $n=1$ of the structure 
function $F_2$. Hence they are the experimentally observable quantity 
and agree with the usual perturbatively defined
distributions only when the higher twists effects are neglected and
the coefficient functions are convoluted. 
Concerning the singlet combination we must regularize the sum rules. Here we have used 
the soft pomeron and regularized the sea quarks distribution 
as $\langle \widetilde{\lambda}_i\rangle = \int_0^1dx \{\lambda_i - ax^{-\alpha_P(0)}\},$ 
where $\alpha_P(0)$ is $\alpha_P(0)=1+b$ and is taken as
$1.0808$.~\cite{lan} The constant $a$ is defined as $\lim_{x\to 0}x^{\alpha_P(0)}\lambda_i =a$
and determined as $a\sim 0.1$ through the sum rule. 
However, this value of $a$ and the regularization dependence
of the regularized quantity should not be taken seriously. 
In the sum rules of the non-singlet combination the regularization dependent terms
drop out due to the fact that the leading  high energy behavior is universal and flavor
symmetric. In this sense we can regard the sum rules of the non-singlet combination
are regularization independent.~\cite{kore95} Further, the perturbative correction 
to this combination is negligible.\\
Now the sum rules for the mean quantum number of the light sea quarks can be obtained
by using the  mean hypercharge sum rule. The results is~\cite{kore95}
\begin{eqnarray}
<Q>_{light\;sea\;quarks}^{proton}
=\frac{1}{3}\int_0^1dx\{ 2\lambda_u - \lambda_d - \lambda_s \}
=\frac{1}{2}[\{\frac{1}{2\pi}P\int_{-\infty}^{\infty}
\frac{d\alpha}{\alpha}A_3(\alpha ,0)-\frac{1}{2}\} \nonumber \\
+\frac{1}{2}\{
\frac{1}{2\pi}\frac{2\sqrt{3}}{3}P\int_{-\infty}^{\infty}\frac{d\alpha}{\alpha}
A_8(\alpha ,0) -1 \} ]
=\frac{1}{6}(I_{\pi} + I_K^p - 2I_K^n)-\frac{1}{2} \sim 0.2,\hspace{1cm}
\end{eqnarray}
where  $I_{\pi}\sim 5.17,I_K^p\sim 2.39,I_K^n\sim 1.61$,\cite{kore93,kore95}
and $A_a(p\cdot x,x^2)$ is a piece of the matrix element of the bilocal currents.
The extension to the ${\bf 8}$ baryon is straightforward by noticing the fact that the
quantity $A_a(\alpha ,0)$ is governed by the chiral $SU(3)\otimes SU(3)$ flavor symmetry 
on the null-plane. We define
$\langle \alpha , p|\frac{1}{2i}
[:\bar{q}(x)\gamma^{\mu}\frac{1}{2}\lambda_a q(0) -
\bar{q}(0)\gamma^{\mu}\frac{1}{2}\lambda_a q(x):]|\beta ,p\rangle_c
=p^{\mu}(A_a(px,x^2))_{\alpha \beta }+x^{\mu}(\bar{A}_a(px,x^2))_{\alpha \beta },$
where $\alpha , \beta $ are the symmetry index 
specifying each member of the ${\bf 8}$ baryon.  
Since the matrix element can be classified by the flavor singlet in the
product ${\bf 8\otimes 8\otimes 8}$, $(A_a(p\cdot x ,0))_{\alpha  \beta }$ 
is decomposed as $(A_a(p\cdot x , 0))_{\alpha  \beta } = if_{\alpha a \beta}F(p\cdot x,0)
+ d_{\alpha  a \beta }D(p\cdot x,0)$ for $a\neq 0$.  Using the value of the modified Gottfried
sum rule and that of the mean hypercharge sum rule, we obtain
$\widetilde{F} \equiv\frac{1}{2\pi} P\int^{\infty}_{-\infty}
\frac{d\alpha}{\alpha}F(\alpha ,0) = 0.89 ,$
$\widetilde{D} \equiv\frac{1}{2\pi} P\int^{\infty}_{-\infty}
\frac{d\alpha}{\alpha}D(\alpha ,0) = -0.50.$
These values give us many relations for the light sea quarks in the ${\bf 8}$ baryon.
The extension to the chiral $SU(N)\otimes SU(N)$ flavor symmetry
is straightforward, and we obtain the result that the heavy sea quark in the ${\bf 8}$ baryon
is universal and abundant.~\cite{kore95}
\section{Soft pion contributions to the strange sea quark}
Now the mean charge sum rule of the light sea quarks in the proton is badly
broken in the phenomenologically determined sea quarks. This is because
the strange sea quark is considered to be suppressed compared
with the non-strange light sea quarks. Here we give one possible scenario
why the strange sea quark becomes large
in the phenomenologically poorly determined small $x$ region.
For this purpose, we first explain the method by Sakai and Yamada.~\cite{sakai}
Let us take the semi-inclusive reaction
$\pi + p \to \pi_{s}(k) + anything$ .  The differential cross section is defined as
$f(k^3,\vec{k}^{\bot},p^0)=k^0\frac{d\sigma}{d^3k},$
where $p^0$ is the energy in the CM frame. Let us assume that at high energy Feynman 
scaling is satisfied as
$f\sim f^{F}(\frac{k^3}{p^0},\vec{k}^{\bot}) + \frac{g(k^3,\vec{k}^{\bot})}{p^0}.$
Then we obtain $\lim_{p^0\to \infty}f^F(\frac{k^3}{p^0},\vec{k}^{\bot}=0)=
f^F(0,0)=\lim_{p^0\to \infty}f(0,0,p^0).$
This means that, in the CM frame, we can regard the pion with its momentum
$k^3<O(p^0)$ and $\vec{k}^{\bot}=0$ as the soft pion.
To examine this, the soft $\pi^-$ production in the deep inelastic inclusive 
lepton-proton scatterings is calculated and compared with the 
the experiment of Harvard-Cornell group~\cite{harvard} under the conditions,
(1)the transverse momentum satisfys $|\vec{k}^{\bot}|^2\leq m_{\pi}^2$ and
(2)the change of $F^-$ can be regarded to be small in the small $x_F$ region.
Here $F^-$ is defined as $F^{-}=\frac{1}{\sigma_T}q^0\frac{d\sigma^{-}}{d^3q}.$
Then, the theoretical value~\cite{kore78} obtained by the light-cone current
algebra~\cite{fritzsch} is about 10\% $\sim$ 20\% of the experimental value.
In the central region, there are many pions from the decay of the resonances, and about
20\% $\sim$ 30\% can be expected to be the pion from the directly produced pion.
Thus the method by Sakai and Yamada may be applicable if we identify the soft pion at high
energy as a directly produced pion. Since the ambiguity comes from the fact
that the pion from a directly produced one has not been separated
from that produced by the decay product of the resonance particles, 
we have calculated a difference of the $\pi^+$ and $\pi^-$.
As far as the production and the decay mechanism is governed by the strong interaction, 
this difference can be expected to be zero in the central region. However, since the soft pion
theorem for the $\pi^+$ and the $\pi^-$ production is charge asymmetric due to pion pole terms, 
if this theorem has any relevance to the high energy reaction,
its remnant remains as a charge asymmetry. Thus, by calculating $F^+ - F^- $, we have found that
the theoretical value is about the same order with the experimental value of the
Harvard-Cornell group. The difference is within a factor 2 at most.~\cite{kore82}\\
Based on this study a soft pion contribution to the Gottfried sum has been calculated.
The phase space of the soft pion is determined as
(1)the transverse momentum satisfys $|\vec{k}^{\bot}|\leq bm_{\pi}$ and
(2)Feynman scaling variable $x_{F}=2k^3/\sqrt{s}$ satisfys $|x_{F}|\leq c$.
we take $b=1$ and $c=0.1$ following the above phenomenological study.
Then we find about $-0.03$ in the Gottfried defect can be explained by this soft pion contribution.
Since the Gottfried defect is about $-0.07$, this corresponds to about $40\%$.
Though the magnitude of the soft pion contribution has a large ambiguity
due to the poorly determined soft pion phase space, it points to a right direction.~\cite{kore20}
Encouraged by this study, the soft pion contribution to the strange sea quark 
has been calculated.~\cite{kore21}\\
In the $SU(4)\otimes SU(4)$ model, we find that the soft pion contribution
to the strange sea quark can be approximately given as
$x\lambda_s|_{soft} = \frac{I_{\pi}}{f_{\pi}^2}[\frac{x}{6}(d_v + u_v)
+ \frac{x}{3}(\lambda_u + \lambda_d) + \frac{3}{4}xg_A^2(0)(1+\langle n \rangle)\lambda_s 
- \frac{5}{3}xg_A(0)(\triangle u_v - \triangle d_v)],$
where we set $\langle n \rangle^{\nu N}=\langle n \rangle^{eN}=\langle n \rangle$,
and $\langle n \rangle$ is the sum of the nucleon and the antinucleon multiplicity.
By denoting $\lambda_s^a = \frac{1}{4}(\lambda_u + \lambda_d)$
as the strange sea quark given 
\epsfysize=4cm \epsfbox{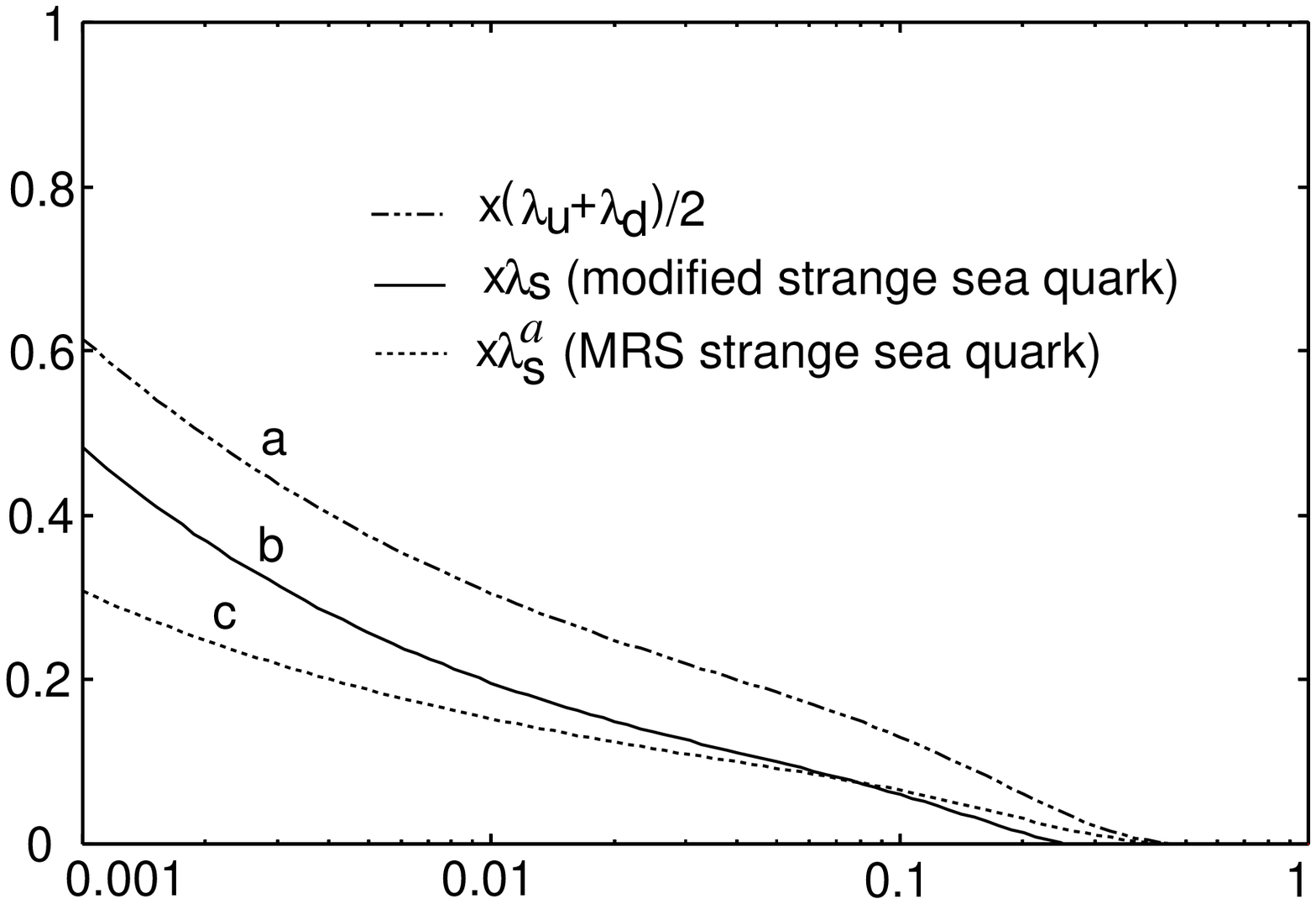}
\begin{minipage}[b]{5.8cm}
by the usual phenomenological analysis,
we define the modified strange sea quark distribution as
$x\lambda_s = x\lambda_s^a + x\lambda_s|_{soft}.$ 
We estimate the soft pion contribution by using the distribution by MRS 
and GS at $Q_0^2=4$GeV$^2$.~\cite{MRS,GS} The result is given in the Figure.
We find that the soft pion contribution is suppressed 
above $x\sim 0.01$, but it becomes very large below it due to the 
\end{minipage}\\
nucleon multiplicity term which comes from the
pion emission pole term where the proper part of the axial-vector current is attached
to the final nucleon. Though the magnitude of the phase
space factor $I_{\pi}$ has a large ambiguity, and the multiple soft pion 
contribution can be expected to become large, the sum rule gives us an adequate quantity.
\section{Conclusion}
The chiral $SU(N)\otimes SU(N)$ flavor symmetry on the null-plane gives us 
the non-pertubative information on sea quarks in the hadron. 
The modified Gottfried sum rule and the mean charge sum rule
of the light sea quarks in the proton are illustrated as examples, 
and a possible contribution from soft pion at high energy to these sum rules is
discussed. The contribution from the soft pion at high energy to the 
Gottfried sum is sizable and can explain a part of the Gottfried defect.
Then, using the same parameters we estimate the contribution to the strange
sea quark  and find that it becomes very large below $x\sim 0.01$ though it is 
greatly suppressed above it.
This fact can explain why the phenomenologically determined strange
sea quark distribution is underestimated in the small $x$ region. We show that,if this 
contribution is added to the ususal strange sea quark distribution 
extrapolated to the small $x$ region,
it can satisfy the mean charge sum rule because the sea quark distribution can become
flavor symmetric in the small $x$ region.~\cite{kore21}

\end{document}